\documentclass[a4paper]{article}
\usepackage[dvipsnames]{xcolor}
\usepackage[utf8]{inputenc}
\usepackage[ngerman,main=english]{babel}
\usepackage{amsmath}
\usepackage{amssymb}
\usepackage{amsthm}
\usepackage{comment}
\usepackage[pdfusetitle]{hyperref}
\hypersetup{
  colorlinks=true,
  allcolors=RoyalBlue,
  pdfborder={0 0 0}
}
\usepackage{orcidlink}
\usepackage{titling}
\usepackage{multicol}
\usepackage{tikz}
\usetikzlibrary{shapes,arrows}
\usepackage[nottoc]{tocbibind}
\usepackage{graphicx}
\usepackage{bbold}
\usepackage{csquotes}

\usepackage{enumitem}
\usepackage[T1]{fontenc}
\usepackage{lmodern}

\usepackage[
  a4paper,
  textwidth=380pt,
  textheight=632pt,
  heightrounded,
  hcentering,
  vcentering
]{geometry}
\usepackage{fancyhdr}
\usepackage{mathrsfs}
\usepackage[normalem]{ulem}

\usepackage{import}

\usepackage[capitalize]{cleveref}

\usepackage[doi=true,isbn=false,url=true,eprint=true,date=year,backend=biber,style=numeric,
  maxbibnames=999]{biblatex} 
\AtEveryBibitem{%
  \iffieldundef{doi}
    {\iffieldundef{eprint}{}{\clearfield{url}}}
    {\clearfield{eprint}\clearfield{url}}}

\renewbibmacro*{journal+issuetitle}{%
  \usebibmacro{journal}%
  \setunit*{\addspace}%
  \iffieldundef{series}
    {}
    {\newunit
     \printfield{series}%
     \setunit{\addspace}}%
  \usebibmacro{volume+number+eid}%
  \setunit{\bibpagespunct}%
  \printfield{pages}%
  \setunit{\addspace}%
  \usebibmacro{issue+date}%
  \setunit{\addcolon\space}%
  \usebibmacro{issue}%
  \newunit}

\renewbibmacro*{note+pages}{%
  \printfield{note}%
  \newunit}

\newcommand{\tr}{\mathrm{tr}}
\newcommand{\e}{\mathrm{e}}
\newcommand{\D}{\mathrm{d}}

\renewcommand{\i}{\mathrm{i}}
\renewcommand{\Phi}{\varPhi}
\renewcommand{\Psi}{\varPsi}
\renewcommand{\Sigma}{\varSigma}

\newcommand{\w}{\omega}

\newcommand{\Z}{\mathbb{Z}}
\newcommand{\C}{\mathbb{C}}

\renewcommand{\H}{\mathscr{H}}
\newcommand{\mA}{\mathcal{A}}

\newcommand{\statement}[1]{\paragraph{#1}\pdfbookmark[1]{#1}{#1}}

\usepackage{mathtools}

\DeclarePairedDelimiter\norm{\lVert}{\rVert}

\usepackage{thmtools}
\usepackage{thm-restate}

\newlength{\thmspacing}
\declaretheoremstyle[
  spaceabove=\thmspacing, spacebelow=\thmspacing,
  headfont=\normalfont\bfseries, bodyfont=\normalfont\itshape
]{thmplain}
\declaretheoremstyle[
  spaceabove=\thmspacing, spacebelow=\thmspacing,
  headfont=\normalfont\bfseries, bodyfont=\normalfont
]{thmdefinition}
\declaretheoremstyle[
  spaceabove=\thmspacing, spacebelow=\thmspacing,
  headfont=\normalfont\itshape, bodyfont=\normalfont
]{thmremark}

\theoremstyle{thmplain}
\newtheorem{theorem}{Theorem}[section]
\newtheorem{lemma}[theorem]{Lemma}

\theoremstyle{thmdefinition}

\theoremstyle{thmremark}

\usepackage[indent=15pt]{parskip}

\usepackage{colonequals}

\newcommand{\scal}[2]{\left\langle #1 , #2 \right\rangle}

\usepackage{microtype}

\pretitle{\centering\normalcolor\bfseries\huge}
\posttitle{\par}

\preauthor{\centering\Large\lineskip .5em\begin{tabular}[t]{c}}
\postauthor{\end{tabular}\par}
\predate{}
\postdate{}
\renewenvironment{abstract}{\quotation}{\endquotation}

\makeatletter
\def\blindfootnote{\gdef\@thefnmark{}\@footnotetext}
\makeatother
\newcommand{\emaillink}[1]{\href{mailto:#1}{#1}}

\title{A Counterexample to the Weak Gibbs and Regularity Conjectures}
\date{}

\author{
    Marius Wesle%
    \texorpdfstring{%
        \,\orcidlink{0009-0006-3365-4037}%
        \thanks{
            \parbox[t]{.8\textwidth}{
                \foreignlanguage{ngerman}{
                    Fachbereich Mathematik,
                    Universität~Tübingen,\newline
                    Auf~der~Morgenstelle~10,
                    72076~Tübingen%
                },
                Germany%
            }
        }
    }{}%
}

\begin{document}

\bgroup
\maketitle\thispagestyle{empty}
\blindfootnote{
    \parbox[t]{.8\textwidth}{
        Email:
        \emaillink{marius.wesle@uni-tuebingen.de}
    }
}
\egroup

\begin{abstract}
  We provide a counterexample to the weak Gibbs and regularity conjectures. The example is a spin-one system on the square lattice with a translation-invariant nearest-neighbor interaction. At high temperatures and high on-site field strengths, it has a unique KMS state that does not have the weak Gibbs or regularity properties.
\end{abstract}

\section{Introduction}

For the finite-volume Gibbs states of a quantum spin system, the relative entropy is equal to the free energy excess multiplied by the inverse temperature, see for example \cite{BratteliRobinson1997}. Schematically, for every density matrix $\sigma$,
\[
  D(\sigma\Vert\gamma)
    =-S(\sigma)+\beta E_H(\sigma)+P_H(\beta),
\]
where $S$, $E_H$, and $P_H$ denote the entropy, energy, and pressure
contribution, respectively, and $\gamma$ is the Gibbs state at inverse
temperature $\beta>0$.
A natural question to ask is whether, for KMS states of infinitely extended systems, the same relation persists for the respective volume-averaged quantities. States that satisfy this property are called regular \cite[Definition~2.15]{JPT24}. In this note, we construct a translation-invariant nearest-neighbor quantum spin system whose unique KMS state is nonregular.

More precisely, consider two translation-invariant states $\nu, \omega$ on the quasi-local algebra $\mA$ of the spin system. Their relative entropy density is defined as
\[
  s(\nu\Vert\omega)
    :=\lim_{Z\nearrow\Z^d}
      \frac{D(\nu_Z\Vert\omega_Z)}{|Z|}
    =\lim_{Z\nearrow\Z^d}
      \frac{\tr_Z\bigl(\nu_Z(\log\nu_Z-\log\omega_Z)\bigr)}{|Z|},
\]
whenever the limit exists. Here $\nu_Z$ and $\omega_Z$ are the local
density matrices, and $Z\nearrow\Z^d$ denotes increasing cubes centered
at the origin.
The state $\omega$ is said to be regular if, for every translation-invariant state $\nu$ on $\mA$, the relative entropy density exists and it holds that
\begin{equation}\label{eq:regularity-identity}
  s(\nu\Vert\omega)=-s(\nu)+\beta e_H(\nu)+p_H(\beta),
\end{equation}
where the entropy density, energy density, and pressure are defined as
\[
  \begin{aligned}
    s(\nu)&=-\lim_{Z\nearrow\Z^d}
      \frac{\tr_Z(\nu_Z\log\nu_Z)}{|Z|},\\
    e_H(\nu)&=\lim_{Z\nearrow\Z^d}
      \frac{\tr_Z(\nu_Z H_Z)}{|Z|},\\
    p_H(\beta)&=\lim_{Z\nearrow\Z^d}
      \frac{\log\tr_Z\exp(-\beta H_Z)}{|Z|}.
  \end{aligned}
\]
A regular state is always KMS \cite[Theorem~2.2 and the remark after Definition~2.15]{JPT24}. Whether the converse of this statement is true is a longstanding open problem. It is explicitly formulated as Conjecture-(R) by \textcite[Section~2.2.6]{JPS26}. The corresponding upper bound on the left-hand side of \eqref{eq:regularity-identity} was established by \textcite{JPT24} using previous results of \textcite{HiaiPetz1993}, and the full identity has been proven for one-dimensional finite-range models and at sufficiently high temperatures \cites{HiaiPetz1993}{HiaiMosonyiOhnoPetz2008}.
In our example, we establish for a KMS state $\omega$ and an explicit translation-invariant state $\nu$ the strict inequality
\begin{equation}\label{eq:introduction-nonregularity}
  \liminf_{Z\nearrow\Z^2}
    \frac{D(\nu_Z\Vert\omega_Z)}{|Z|}
    <-s(\nu)+\beta e_H(\nu)+p_H(\beta),
\end{equation}
thereby disproving the regularity conjecture.

Quantum weak Gibbsianity is a stronger property that implies regularity \cite[Propositions~2.10 and~2.14]{JPT24} and is of independent interest.
A translation-invariant state $\omega$ is said to be weak Gibbs at inverse temperature $\beta>0$ if there are constants $c_Z\geq0$ for finite lattice subsets $Z$ such that
\[
  \e^{-c_Z}\frac{\exp(-\beta H_Z)}{\tr_Z\exp(-\beta H_Z)}
  \leq\omega_Z
  \leq\e^{c_Z}\frac{\exp(-\beta H_Z)}{\tr_Z\exp(-\beta H_Z)},
  \qquad
  \lim_{Z\nearrow\Z^d}\frac{c_Z}{|Z|}=0.
\]
The quantum definition was formalized in \cite{JPT24}, following earlier quantum Gibbs comparison estimates \cite{ArakiIon1974,HiaiPetz1993} and the classical notion of weak Gibbs measures \cite{Yuri2002}.
Since weak Gibbsianity implies regularity, a weak Gibbs state is always a KMS state, and it was conjectured in \cite{JPT24Adiabatic} that the converse is also true. We refer to this as the weak Gibbs conjecture.
Weak Gibbsianity of KMS states has been established in finite-range one-dimensional systems and at sufficiently high temperature \cite{JPT24}, and for gauge-invariant free lattice fermions \cite{JaksicPilletSzczepanek2026Fermions}.
Since the KMS state in our example is nonregular, it is also not weak Gibbs.

Pettinari \cite{Pettinari2026} recently disproved the weak Gibbs
conjecture for a translation-invariant nearest-neighbor qubit interaction
on $\Z^2$. The present work was developed independently and establishes
failure of regularity in a model with a unique KMS state. Both proofs
build on the multiscale branching idea of Bouch \cite{Bouch15}.

Our example is a spin-one system on the square lattice with nearest-neighbor interaction terms and a uniform on-site field. It is closely related to models studied in the context of quantum breakdown \cite{Lian2023,HuHanLian2026}. Its precise definition can be found in Section~\ref{sec:model}. It has three important features:
First, the Hamiltonian preserves an exponentially weighted occupation number, which we use to express the relative entropy of the empty product state in terms of an empty-region probability.
Second, the interaction terms can amplify the occupation number in a cascading manner, allowing for configurations that are entirely localized within a boundary region to have volume-order effects. Third, the on-site field provides a parameter with which we can obtain nonregularity while retaining uniqueness of the equilibrium state.

Our proof proceeds as follows.
We first prove uniqueness for small inverse temperature, with a threshold independent of the field. For this, we use the results of \textcite{RouzeStilckFranca2026}, which provide, for each finite volume, an auxiliary dissipative dynamics that converges to the Gibbs state, called a Gibbs sampler. We construct an infinite-volume version of this Gibbs sampler that sends all observables to scalar multiples of the identity and show that all KMS states are invariant under this evolution, from which we can then conclude uniqueness. This gives us a $\beta_0>0$ such that, for all $0<\beta<\beta_0$ and all field strengths $h$, there is a unique KMS state $\w$. Note that choosing $h$ large moves us outside the usual high-temperature regime, where regularity is known to hold.
We test regularity against the empty product state $\nu$. Conservation of the exponentially weighted occupation makes the empty vector an eigenvector of each local density matrix of $\w$, so the relative entropy is minus the logarithm of the empty-region probability. On each box $\Lambda_L$ of side length $2L$ around the origin, we have
\[
  D(\nu_{\Lambda_L}\Vert\omega_{\Lambda_L})
    =-\log\omega(\nu_{\Lambda_L}).
\]
We then use positivity of matrix entries and bounds on the
partition function to obtain a bound on the above expression in terms of the local partition function $\mathcal Z_{\Lambda_L}$, an imaginary-time return amplitude $G_{\Lambda_L^{+}}=\scal{u_{\Lambda_L^{+}}}{\exp(-\beta H_{\Lambda_L^{+}})u_{\Lambda_L^{+}}}$, and a comparison error term. The vector $u_{\Lambda_L^{+}}$ in this return amplitude is defined on $\Lambda_L^{+}$, an enlarged box with one additional row and column to the bottom and left. It is empty on $\Lambda_L$ with only the additional boundary occupied. This allows us to obtain a comparison error of boundary order:
\[
  D(\nu_{\Lambda_L}\Vert\omega_{\Lambda_L})
    \leq\log\mathcal Z_{\Lambda_L}-\log G_{\Lambda_L^{+}}+O(L).
\]
Thus, we want to show that $-\log G_{\Lambda_L^{+}}\leq \bigl(-s(\nu)+\beta e_H(\nu)-\delta\bigr)|\Lambda_L|+O(L)$ for some $\delta>0$. We show this along a subsequence of boxes with $L=2^k-1$, which yields \eqref{eq:introduction-nonregularity}. For the empty product state $\nu$, we have $s(\nu)=0$ and $e_H(\nu)= 2h$. Indeed, we can show that $-\log G_{\Lambda_L^{+}}\leq\bigl(\frac{3\beta h}{2}+2\log(1+h)+14\log2\bigr)|\Lambda_L|+O(L)$. Choosing $h$ large enough completes the proof.
The final estimate is obtained by recursively bounding the return amplitude from below along shrinking dyadic boxes. This is where the special form of the interactions and the multiscale branching idea of Bouch enter.

Section~\ref{sec:model} gives the precise description of the model and states the main result.
Section~\ref{sec:main-proof} proves the theorem from three lemmas:
uniqueness of the KMS state, a comparison between the empty-region probability and
the return amplitude, and the recursive return-amplitude estimate. Their proofs
are given in Sections~\ref{sec:uniqueness-proof},
\ref{sec:probability-proof}, and~\ref{sec:return-proof}, respectively.

\section{Model and main result}\label{sec:model}

We define a quantum spin system on $\Z^2$. The local Hilbert space is $\C^3$ with basis $|0\rangle,|1\rangle,|2\rangle$. For a finite subset $Z \subseteq \Z^2$,
the local Hilbert space and the algebra of local observables are
\[
  \H_Z=\bigotimes_{x\in Z}\C^3,
  \qquad
  \mA_Z=\bigotimes_{x\in Z}\mathcal B(\C^3).
\]
We use the canonical embeddings obtained by tensoring with the identity
and write $\mA$ for the usual quasi-local C$^*$-algebra obtained by norm completion of the union of all local algebras.
For a state $\rho$ on $\mA$, we denote by $\rho_Z\in \mA_Z$ the unique density matrix with $\rho(A)=\tr_Z(\rho_Z A)$ for $A\in\mA_Z$.
The relative entropy is defined as
\[D(\rho_Z\Vert\sigma_Z)=\tr_Z\bigl(\rho_Z(\log\rho_Z-\log\sigma_Z)\bigr),\] with
value $+\infty$ if
$\operatorname{supp}\rho_Z\nsubseteq\operatorname{supp}\sigma_Z$. Here $\log$ denotes the natural logarithm.

On the local Hilbert space, define the particle number and annihilation operators
\[
  n=|1\rangle\langle1|+2|2\rangle\langle2|,
  \qquad c=|0\rangle\langle1|+|1\rangle\langle2|
\]
and denote by $n_x$ and $c_x$ their embeddings into $\mA$ acting on site $x\in \Z^2$.
We write $x\to y$ if $y-x$ is one of the two positive coordinate vectors.
For a field $h\geq0$, the Hamiltonian on a finite lattice subset $Z$ is
\begin{equation}\label{eq:spin-one-Hamiltonian}
  H_Z=h\sum_{x\in Z}(2-n_x)
      -\sum_{\substack{x\to y\\x,y\in Z}}F_{xy},
  \qquad
  F_{xy}=c_x(c_y^*)^2+c_x^*c_y^2.
\end{equation}
On each directed bond, $F_{xy}$ is a hopping term that turns one particle into two along the bond and two into one in the opposite direction. It is of norm one and, on any local Hilbert space $\H_Z$ with $x,y\in Z$, its occupation-basis matrix entries are
nonnegative.
For $\beta>0$, the local partition function and Gibbs density matrix are
\[
  \mathcal Z_Z=\tr_Z\exp(-\beta H_Z),
  \qquad
  \gamma_Z=\frac{\exp(-\beta H_Z)}{\mathcal Z_Z}.
\]

The Hamiltonian conserves the dyadically weighted particle number
\begin{equation}\label{eq:conserved-charge}
  C_Z=\sum_{x\in Z}2^{-(x_1+x_2)}n_x.
\end{equation}
Indeed, for every $x\in Z$, we have $[n_x,C_Z]=0$, and for $x\to y$ with $x,y\in Z$,
using $[n,c]=-c$ and $[n,c^*]=c^*$, we also obtain
\[
  [F_{xy},C_Z]
    =\bigl(2^{-(x_1+x_2)}-2\cdot2^{-(y_1+y_2)}\bigr)
      \bigl(c_x(c_y^*)^2-c_x^*c_y^2\bigr)=0,
\]
since $2^{-(y_1+y_2)}=\tfrac12 2^{-(x_1+x_2)}$. This gives us
\begin{equation}\label{eq:charge-conservation}
  [H_Z,C_Z]=0,
  \qquad [\gamma_Z,C_Z]=0.
\end{equation}

\begin{theorem}\label{thrm:result}
  There exists $\beta_0>0$ such that, for all $0<\beta<\beta_0$ and $h\geq0$, the
  translation-invariant nearest-neighbor Hamiltonian
  \eqref{eq:spin-one-Hamiltonian} has exactly one $\beta$-KMS state
  on $\mA$. There exists an $h_0$, dependent on $\beta$, such that, for all $h>h_0$, this state is not regular and hence not weak Gibbs.
\end{theorem}

\section{Proof of \texorpdfstring{\hyperref[thrm:result]{Theorem~\ref*{thrm:result}}}{Theorem~\ref*{thrm:result}}}\label{sec:main-proof}

We carry out the proof as outlined in the introduction using three lemmas whose proofs are postponed to the following sections.
We first obtain a unique KMS state from the following uniqueness lemma. Since it is a limit of local Gibbs states, we will be able to estimate its local density matrices by finite-volume comparisons.

\begin{lemma}\label{lem:kms-uniqueness}
  There exists $\beta_0>0$ such that, for all $0<\beta<\beta_0$ and $h\geq0$,
  the Hamiltonian
  \eqref{eq:spin-one-Hamiltonian} has a unique $\beta$-KMS state. This state is translation invariant and is the local limit
  of the finite-volume Gibbs states along growing boxes.
  \par\nobreak\noindent{\normalfont[\hyperref[sec:uniqueness-proof]{Proof}]}
\end{lemma}

Fix $0<\beta<\beta_0$. For now, let $h\geq0$ be arbitrary and
let $\omega$ be the state supplied by Lemma~\ref{lem:kms-uniqueness}.
This setup allows us to choose the field after deriving our relative-entropy bound while retaining uniqueness.
We will bound the relative entropy between $\w$ and the empty product state $\nu$, whose local density matrices are
\[
  \nu_Z=\bigotimes_{x\in Z}|0\rangle\langle0|_x.
\]
Every bond interaction term annihilates the empty vector, so
\[
  s(\nu)=0,
  \qquad \tr_Z\bigl(\nu_Z H_Z\bigr)=2h|Z|,
  \qquad e_H(\nu)=2h.
\]
Regularity of $\omega$ would require
\[
  \lim_{Z\nearrow\Z^2}\frac{D(\nu_Z\Vert\omega_Z)}{|Z|}
    =2\beta h+p_H(\beta).
\]
We will obtain a strictly smaller upper bound along a sequence of squares.

For integers $L\geq0$, set
\[
  \Lambda_L=\{-L,\ldots,L\}^2.
\]
Fix $L$ and let $A\in\mA_{\Lambda_L}$ be arbitrary. We have
\[
  \begin{aligned}
    \tr_{\Lambda_L}\bigl([\omega_{\Lambda_L},C_{\Lambda_L}]A\bigr)
      &=\omega([C_{\Lambda_L},A])\\
      &=\lim_{K\to\infty}\tr_{\Lambda_K}
          \bigl(\gamma_{\Lambda_K}[C_{\Lambda_L},A]\bigr)\\
      &=\lim_{K\to\infty}\tr_{\Lambda_K}
          \bigl(\gamma_{\Lambda_K}[C_{\Lambda_K},A]\bigr)\\
      &=\lim_{K\to\infty}\tr_{\Lambda_K}
          \bigl([\gamma_{\Lambda_K},C_{\Lambda_K}]A\bigr)=0.
  \end{aligned}
\]
Here we used cyclicity of the trace and local convergence from
Lemma~\ref{lem:kms-uniqueness}. For $K>L$, the commutators
$[C_{\Lambda_L},A]$ and $[C_{\Lambda_K},A]$ agree because the terms
of $C_{\Lambda_K}$ outside $\Lambda_L$ commute with $A$.
The last equality follows from \eqref{eq:charge-conservation}.
Because this holds for every $A\in\mA_{\Lambda_L}$, we have
$[\omega_{\Lambda_L},C_{\Lambda_L}]=0$.
All weights in \eqref{eq:conserved-charge} are positive,
so the zero eigenspace of $C_{\Lambda_L}$ consists precisely of the
multiples of the empty vector. Thus $\nu_{\Lambda_L}$ is its spectral
projection at zero and also commutes with $\omega_{\Lambda_L}$.
Since $\nu_{\Lambda_L}$ has rank one, it follows that
\[
  \omega_{\Lambda_L}\nu_{\Lambda_L}
    =\omega(\nu_{\Lambda_L})\nu_{\Lambda_L}.
\]
If $\omega(\nu_{\Lambda_L})>0$, the rank-one projection
$\nu_{\Lambda_L}$ has zero entropy, and functional calculus gives
$(\log\omega_{\Lambda_L})\nu_{\Lambda_L}
  =\log\omega(\nu_{\Lambda_L})\nu_{\Lambda_L}$.
Hence
\begin{equation}\label{eq:entropy-empty-probability}
  \begin{aligned}
    D(\nu_{\Lambda_L}\Vert\omega_{\Lambda_L})
      &=\tr_{\Lambda_L}\bigl(\nu_{\Lambda_L}
          (\log\nu_{\Lambda_L}-\log\omega_{\Lambda_L})\bigr)\\
      &=-\tr_{\Lambda_L}\bigl(\nu_{\Lambda_L}\log\omega_{\Lambda_L}\bigr)\\
      &=-\log\omega(\nu_{\Lambda_L}).
  \end{aligned}
\end{equation}
If $\omega(\nu_{\Lambda_L})=0$, the support of $\nu_{\Lambda_L}$
is not contained in that of $\omega_{\Lambda_L}$, so the relative
entropy is $+\infty$. Thus \eqref{eq:entropy-empty-probability} also
holds with the convention $-\log0=+\infty$.

\begin{figure}[htbp]
  \centering
  \begin{tikzpicture}[x=0.46cm,y=0.46cm]
    \foreach \shift in {0,8,16} {
      \begin{scope}[shift={(\shift,0)}]
        \foreach \i in {-3,...,2} {
          \foreach \j in {-3,...,2} {
            \fill[black!20] (\i,\j) circle[radius=1.3pt];
          }
        }
      \end{scope}
    }

    \filldraw[fill=blue!8,draw=black,thick]
      (-2.45,-2.45) rectangle (2.45,2.45);
    \foreach \i in {-2,...,2} {
      \foreach \j in {-2,...,2} {
        \fill (\i,\j) circle[radius=1.3pt];
      }
    }
    \draw (0,0) circle[radius=2.8pt];
    \node[anchor=south west,inner sep=1pt] at (0.2,0.15) {$0$};
    \node at (0,3.25) {$\Lambda_L$};
    \node[font=\small] at (-2,-4.25) {$-L$};
    \node[font=\small] at (2,-4.25) {$L$};
    \draw[black!45] (-2,-2.55) -- (-2,-3.75);
    \draw[black!45] (2,-2.55) -- (2,-3.75);

    \begin{scope}[shift={(8,0)}]
      \filldraw[fill=orange!20,draw=black,thick]
        (-3.45,-3.45) -- (2.45,-3.45) -- (2.45,-2.45)
        -- (-2.45,-2.45) -- (-2.45,2.45) -- (-3.45,2.45) -- cycle;
      \foreach \i in {-3,...,2} {
        \fill (\i,-3) circle[radius=1.3pt];
      }
      \foreach \j in {-2,...,2} {
        \fill (-3,\j) circle[radius=1.3pt];
      }
      \node at (-0.5,3.25) {$B_L$};
      \node[font=\small] at (-0.5,-4.25) {added sites};
    \end{scope}

    \begin{scope}[shift={(16,0)}]
      \fill[blue!8] (-2.45,-2.45) rectangle (2.45,2.45);
      \fill[orange!20]
        (-3.45,-3.45) -- (2.45,-3.45) -- (2.45,-2.45)
        -- (-2.45,-2.45) -- (-2.45,2.45) -- (-3.45,2.45) -- cycle;
      \draw[thick] (-3.45,-3.45) rectangle (2.45,2.45);
      \draw[dashed] (-2.45,2.45) -- (-2.45,-2.45) -- (2.45,-2.45);
      \foreach \i in {-3,...,2} {
        \foreach \j in {-3,...,2} {
          \fill (\i,\j) circle[radius=1.3pt];
        }
      }
      \draw (0,0) circle[radius=2.8pt];
      \node[anchor=south west,inner sep=1pt] at (0.2,0.15) {$0$};
      \node at (-0.5,3.25) {$\Lambda_L^{+}$};
      \node[font=\small] at (-3,-4.25) {$-L-1$};
      \node[font=\small] at (2,-4.25) {$L$};
      \draw[black!45] (-3,-3.55) -- (-3,-3.75);
      \draw[black!45] (2,-3.55) -- (2,-3.75);
    \end{scope}
    \node at (3.5,-0.5) {$\cup$};
    \node at (11.5,-0.5) {$=$};
  \end{tikzpicture}
  \caption{Starting with $\Lambda_L$ centered at the origin, add the
    bottom row and left column $B_L$ to obtain
    $\Lambda_L^{+}=\Lambda_L\cup B_L$.
    }
  \label{fig:regions}
\end{figure}
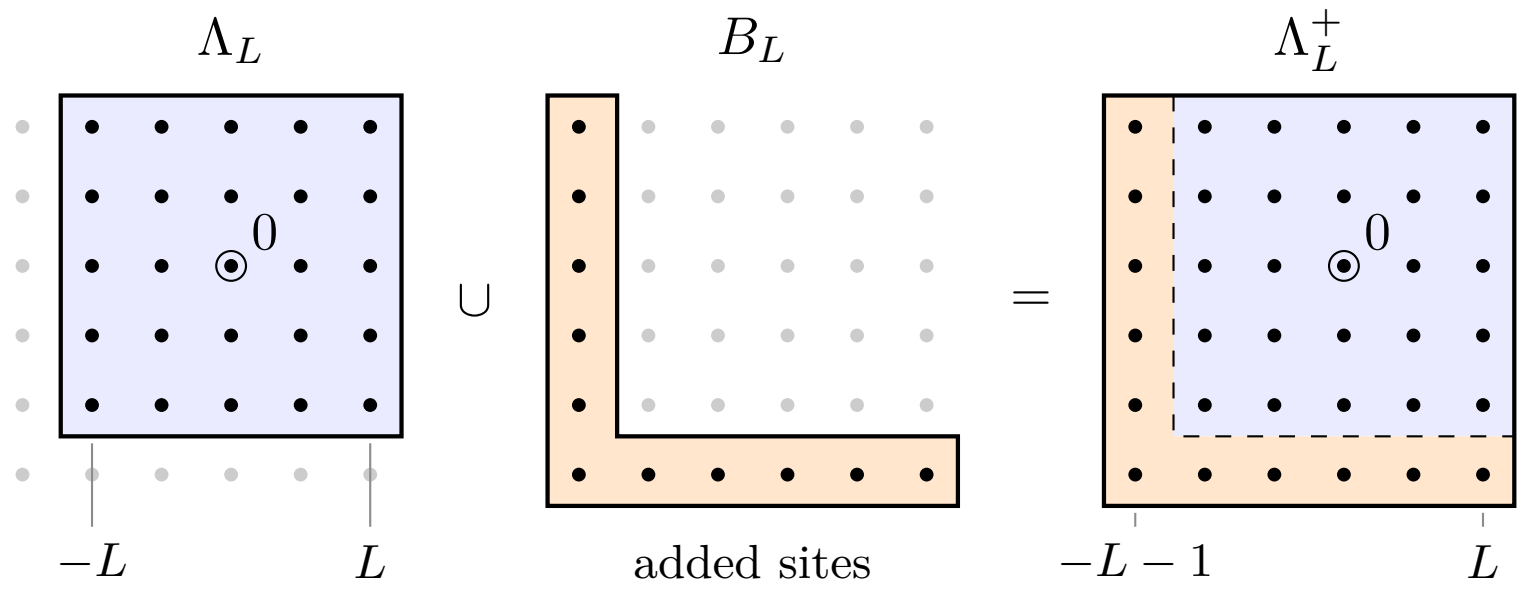

It therefore suffices to bound the empty-region probability $\omega(\nu_{\Lambda_L})$ from below.
We compare it with an imaginary-time return amplitude that allows particles to enter
the square from its bottom and left edges. Define the enlarged square
and the added boundary by
\[
  \Lambda_L^{+}=\{-L-1,\ldots,L\}^2,
  \qquad
  B_L=\Lambda_L^{+}\setminus\Lambda_L.
\]
Thus $B_L$ consists of the added bottom and left edges, and
$|B_L|=4L+3$; see Figure~\ref{fig:regions}.

Let $u_{\Lambda_L^{+}}$ be the product vector with occupation zero on
$\Lambda_L$ and occupation two on $B_L$,
\[
  u_{\Lambda_L^{+}}
    =\left(\bigotimes_{x\in\Lambda_L}|0\rangle_x\right)
       \otimes\left(\bigotimes_{x\in B_L}|2\rangle_x\right).
\]
Define the return amplitude by
\[
  G_{\Lambda_L^{+}}
    =\scal{u_{\Lambda_L^{+}}}{\exp(-\beta H_{\Lambda_L^{+}})u_{\Lambda_L^{+}}}.
\]
The boundary spins participate in the evolution, with their occupations
prescribed only in the initial and final vector $u_{\Lambda_L^{+}}$.
The following lemma bounds the empty-region probability in terms of
this return amplitude, with a correction proportional to $|B_L|$.

\begin{lemma}\label{lem:probability-return}
  For the unique $\beta$-KMS state $\omega$ fixed above and every
  $L\geq1$,
  \begin{equation}\label{eq:probability-return-bound}
    \omega(\nu_{\Lambda_L})
      \geq\e^{-(4\beta+\log3)|B_L|}
          \frac{G_{\Lambda_L^{+}}}{\mathcal Z_{\Lambda_L}}.
  \end{equation}
  \par\nobreak\noindent{\normalfont[\hyperref[sec:probability-proof]{Proof}]}
\end{lemma}

Since $G_{\Lambda_L^{+}}>0$, this also shows that
$\omega(\nu_{\Lambda_L})>0$. Taking minus the logarithm and using
\eqref{eq:entropy-empty-probability} gives
\begin{equation}\label{eq:entropy-return-bound}
  D(\nu_{\Lambda_L}\Vert\omega_{\Lambda_L})
    \leq\log\mathcal Z_{\Lambda_L}-\log G_{\Lambda_L^{+}}
        +(4\beta+\log3)|B_L|.
\end{equation}
Since $|B_L|/|\Lambda_L|\to0$ and
$\log\mathcal Z_{\Lambda_L}/|\Lambda_L|\to p_H(\beta)$, it remains
to bound $-\log G_{\Lambda_L^{+}}/|\Lambda_L|$ strictly below
$2\beta h$ along a sequence of squares. The next lemma provides the required
return bound by repeatedly subdividing a square into four smaller ones.

\begin{lemma}\label{lem:return-amplitude}
  Set $L_k=2^k-1$ for integers $k\geq0$.
  For every $\beta>0$, $h\geq0$, and integer $k\geq1$,
  \begin{equation}\label{eq:return-amplitude-lower-bound}
    -\log G_{\Lambda_{L_k}^{+}}
      \leq\left(\frac{3\beta h}{2}+2\log(1+h)+14\log2\right)
        |\Lambda_{L_k}^{+}|.
  \end{equation}
  \par\nobreak\noindent{\normalfont[\hyperref[sec:return-proof]{Proof}]}
\end{lemma}

We can now choose the field. With $\beta$
fixed as above, choose an $h>0$ so large that
\[
  \delta:=\frac{\beta h}{2}-2\log(1+h)-14\log2>0.
\]
By Lemma~\ref{lem:return-amplitude}, for every $k\geq1$,
\[
  -\log G_{\Lambda_{L_k}^{+}}
    \leq(2\beta h-\delta)|\Lambda_{L_k}^{+}|.
\]
Substituting this bound into \eqref{eq:entropy-return-bound} gives
\begin{align}
  D(\nu_{\Lambda_{L_k}}\Vert\omega_{\Lambda_{L_k}})
    &\leq\log\mathcal Z_{\Lambda_{L_k}}
      +(2\beta h-\delta)|\Lambda_{L_k}^{+}|
      +(4\beta+\log3)|B_{L_k}| \notag\\
    &=\log\mathcal Z_{\Lambda_{L_k}}
      +(2\beta h-\delta)|\Lambda_{L_k}| \notag\\
    &\qquad +(2\beta h-\delta+4\beta+\log3)|B_{L_k}|.
    \label{eq:finite-volume-deficit}
\end{align}
Here we used $|\Lambda_{L_k}^{+}|=|\Lambda_{L_k}|+|B_{L_k}|$.
Thus the relative-entropy deficit is of volume order, while the
error is of boundary order.

Dividing \eqref{eq:finite-volume-deficit} by $|\Lambda_{L_k}|$ and taking the limsup yields
\begin{equation}\label{eq:regularity-deficit}
\limsup_{k\to\infty}
  \frac{D(\nu_{\Lambda_{L_k}}\Vert\omega_{\Lambda_{L_k}})}
       {|\Lambda_{L_k}|}
  \leq p_H(\beta)+2\beta h-\delta
  <p_H(\beta)+2\beta h.
\end{equation}
Since $\nu$ is translation invariant, regularity of $\omega$ would
instead give
\[
  \lim_{L\to\infty}
    \frac{D(\nu_{\Lambda_L}\Vert\omega_{\Lambda_L})}{|\Lambda_L|}
    =-s(\nu)+\beta e_H(\nu)+p_H(\beta)
    =2\beta h+p_H(\beta).
\]
The same limit would hold along $L=L_k$, contradicting
\eqref{eq:regularity-deficit}. Hence $\omega$ is not regular.

\section{Proof of \texorpdfstring{\hyperref[lem:kms-uniqueness]{Lemma~\ref*{lem:kms-uniqueness}}}{Lemma~\ref*{lem:kms-uniqueness}}}\label{sec:uniqueness-proof}

The proof uses an auxiliary evolution of observables, called a Gibbs
sampler. In a finite volume, this evolution preserves the Gibbs
expectation and sends each observable to that expectation times the
identity in the limit $t\to\infty$. We construct its infinite-volume limit and show that, as $t\to\infty$, it
sends every local observable to a scalar multiple of the identity given by the expectation in a KMS-state $\omega$.
We then prove that it preserves every KMS state. These properties
force all KMS states to agree with $\omega$. Indeed, if $\mathcal T_t(A)$ tends to
$\omega(A)\mathbb1$ and $\varphi$ is invariant under $\mathcal T_t$, then
$\varphi(A)=\varphi(\mathcal T_t(A))\to \omega(A)$.

We write $\mathcal T_t$ for this auxiliary evolution and $\tau_t$
for the physical time evolution entering the KMS condition.

\paragraph*{1. The finite-volume Gibbs sampler.}
We use the following consequence of
\cite[Theorems~6.1 and~4.4]{RouzeStilckFranca2026}.
For spins with local Hilbert space $\C^3$ on the square lattice,
there is an $\varepsilon_0\in(0,1]$ such that, at inverse temperature one,
an arbitrary sum of on-site Hamiltonians may be perturbed by
nearest-neighbor terms of total norm less than $\varepsilon_0$ at
each site, and the resulting Hamiltonian admits the construction and
estimates stated below. The threshold and convergence
rate are independent of the on-site Hamiltonians and the volume.
We apply this result to $\beta H_Z$, whose Gibbs matrix at inverse
temperature one is $\gamma_Z$. Its on-site terms are
$\beta h(2-n_x)$, while
\[
  \sup_x\sum_{\substack{y\to z\\x\in\{y,z\}}}
    \norm{\beta F_{yz}}=4\beta.
\]
Thus we may take $\beta_0=\varepsilon_0/4$, independently of $h$.
Fix $0<\beta<\beta_0$ and $h\geq0$ for the rest of the proof.

For every finite $Z$ and $x\in Z$, the construction supplies a
unital completely positive map $\Phi_{x,Z}$ on $\mA_Z$ satisfying
\[
  \tr_Z\bigl(\gamma_Z\Phi_{x,Z}(A)\bigr)=\tr_Z(\gamma_Z A).
\]
We extend these maps to $\mA$ by the identity map outside $Z$ and
use the operator norm for maps on $\mA$. Set
\[
  \mathcal T_{Z,t}
    =\exp\!\left[t\sum_{x\in Z}(\Phi_{x,Z}-\mathrm{id})\right].
\]
The cited theorems give, for self-adjoint $A\in\mA_Z$ and $t\geq0$,
\begin{equation}\label{eq:sampler-mixing}
  \norm{\mathcal T_{Z,t}(A)-\tr_Z(\gamma_Z A)\mathbb1}
    \leq2|Z|\e^{-\kappa t}\norm{A},
    \qquad \kappa=\frac{1-\e^{-1}}2.
\end{equation}
By \cite[Lemma~B.1]{RouzeStilckFranca2026}, there are $b,\mu>0$,
independent of $h$ and $Z$, such that
\begin{equation}\label{eq:sampler-locality}
  \norm{\Phi_{x,Z}-\Phi_{x,Z\cap(x+\Lambda_\ell)}}
    \leq b\e^{-\mu\ell},\qquad \ell\geq1.
\end{equation}

For the KMS argument, we also need the coefficients defining these
updates. By \cite[Proposition~3.4]{RouzeStilckFranca2026},
\[
  \begin{aligned}
    \Phi_{x,Z}&=\exp(\mathcal K_{x,Z}),\\
    \mathcal K_{x,Z}(A)
      &=\sum_{a=1}^9L_{x,a,Z}^*AL_{x,a,Z}+D_{x,Z}A+AD_{x,Z}^*,
  \end{aligned}
\]
where $L_{x,a,Z},D_{x,Z}\in\mA_Z$. With $Q_Z=\exp(-\beta H_Z/2)$, they satisfy
\begin{equation}\label{eq:sampler-balance}
  \begin{aligned}
    &L_{x,a,Z}Q_Z=Q_ZL_{x,a,Z}^*,
      \qquad Q_ZD_{x,Z}=D_{x,Z}^*Q_Z,\\
    &\sum_{a=1}^9L_{x,a,Z}^*L_{x,a,Z}+D_{x,Z}+D_{x,Z}^*=0.
  \end{aligned}
\end{equation}
Finally, the estimates in the proof of
\cite[Lemma~6.3]{RouzeStilckFranca2026} give, with
$Z_{x,\ell}=Z\cap(x+\Lambda_\ell)$,
\begin{equation}\label{eq:sampler-coefficient-locality}
  \begin{aligned}
    &\max_{1\leq a\leq9}
      \norm{L_{x,a,Z_{x,\ell+1}}-L_{x,a,Z_{x,\ell}}}
      +\norm{D_{x,Z_{x,\ell+1}}-D_{x,Z_{x,\ell}}}\\
    &\qquad\leq C_0(\ell+1)^3\e^{-\mu\ell},\qquad \ell\geq1,
  \end{aligned}
\end{equation}
uniformly in $h$ and $Z$.

\paragraph*{2. The infinite-volume sampler and its convergence.}
We first note that $\mathcal T_{Z,t}$ is a unital completely
positive contraction. Indeed,
\[
  \mathcal T_{Z,t}
    =\e^{-|Z|t}\sum_{n=0}^{\infty}\frac{t^n}{n!}
      \sum_{x_1,\ldots,x_n\in Z}
        \Phi_{x_1,Z}\cdots\Phi_{x_n,Z},
\]
where the $n=0$ term is the identity map. This is a convex
combination of compositions of unital completely positive maps.

We next take limits of the individual updates. Telescoping
\eqref{eq:sampler-coefficient-locality} gives
\[
  \norm{D_{x,Z}-D_{x,Z_{x,\ell}}}
    \leq C_0\sum_{r=\ell}^{\infty}(r+1)^3\e^{-\mu r},
\]
and the same bound holds for each $L_{x,a,Z}$. The tail tends to
zero uniformly in $Z$, and once $Z$ contains $x+\Lambda_\ell$,
the restricted coefficients depend only on that square. The
coefficients are therefore Cauchy, with norm limits $L_{x,a}$ and
$D_x$. Hence the generators converge in norm to
\[
  \mathcal K_x(A)=\sum_{a=1}^9L_{x,a}^*AL_{x,a}+D_xA+AD_x^*.
\]
Continuity of the exponential gives the norm limit
$\Phi_x=\exp(\mathcal K_x)$ of $\Phi_{x,Z}$. Passing to the limit
in the last identity of \eqref{eq:sampler-balance} and in
\eqref{eq:sampler-locality} gives
\begin{equation}\label{eq:sampler-limiting-identities}
  \sum_{a=1}^9L_{x,a}^*L_{x,a}+D_x+D_x^*=0,
  \qquad
  \norm{\Phi_x-\Phi_{x,x+\Lambda_\ell}}\leq b\e^{-\mu\ell}.
\end{equation}

Fix $L\geq1$ and $A\in\mA_{\Lambda_L}$. To prove that the
finite-volume evolutions are Cauchy, we compare them with a common
restricted evolution. For integers $K\geq4L$ and
$1\leq\ell\leq K/4$, set
\[
  \mathcal S_{\Lambda_K,\ell,t}
    =\exp\!\left[t\sum_{x\in\Lambda_K}
      (\Phi_{x,x+\Lambda_\ell}-\mathrm{id})\right].
\]
These updates act inside $\Lambda_{\lfloor3K/2\rfloor}$,
so $\mathcal S_{\Lambda_K,\ell,s}(A)$ remains supported there for every
$s\geq0$.

Let $J\geq2K$. Adding and subtracting the updates of radius
$\ell$ centered in $\Lambda_{2K}$, Duhamel's formula gives
\[
  \begin{aligned}
    &\mathcal T_{\Lambda_J,t}(A)-\mathcal S_{\Lambda_K,\ell,t}(A)\\
    &\quad=\int_0^t\mathcal T_{\Lambda_J,t-s}\left[
      \begin{aligned}
      &\sum_{x\in\Lambda_{2K}}
        \bigl(\Phi_{x,\Lambda_J}
          -\Phi_{x,\Lambda_J\cap(x+\Lambda_\ell)}\bigr)\\
      &+\sum_{x\in\Lambda_{2K}\setminus\Lambda_K}
        \bigl(\Phi_{x,\Lambda_J\cap(x+\Lambda_\ell)}
          -\mathrm{id}\bigr)\\
      &+\sum_{x\in\Lambda_J\setminus\Lambda_{2K}}
        (\Phi_{x,\Lambda_J}-\mathrm{id})
      \end{aligned}
      \right]\mathcal S_{\Lambda_K,\ell,s}(A)\,\D s.
  \end{aligned}
\]
We bound the action of each sum on $\mathcal S_{\Lambda_K,\ell,s}(A)$.

In the first sum, each difference has norm at most
$b\e^{-\mu\ell}$ by \eqref{eq:sampler-locality}. Contractivity
therefore bounds its contribution by
$b|\Lambda_{2K}|\e^{-\mu\ell}\norm{A}$.

For the second sum, we use that $\mathcal S_{\Lambda_K,\ell,s}$ is generated
by uniformly bounded maps of range $2\ell$ in the $\ell^\infty$-metric, each generating
contractions. The maps in the sum annihilate $\mathbb1$ and are
supported at distance at least $K-L-\ell\geq K/2$ from
$\Lambda_L$. A finite-range Lieb--Robinson bound therefore controls
their action on the evolved observable $\mathcal S_{\Lambda_K,\ell,s}(A)$.
Applying \cite[Theorem~3.5]{RubilianiWesleWessel2026}, using
their norm bound of two, and summing over the centers gives
\[
  2|\Lambda_{2K}||\Lambda_L|
    \exp\!\left(C\ell^4s-\frac{K}{4\ell}\right)\norm{A}.
\]
Here $C$ is independent of $h$, $J$, $K$, and $\ell$.

In the third sum, the centers lie outside $\Lambda_{2K}$, while
$\mathcal S_{\Lambda_K,\ell,s}(A)$ stays supported in the smaller box
$\Lambda_{\lfloor3K/2\rfloor}$. We restrict each update to a square
that misses this support. The restricted update leaves the observable
unchanged, so only the restriction error contributes.
For a center with $|x|_\infty=m>2K$, choose the restriction radius
$\lfloor m/4\rfloor$. Every site $y$ in this square satisfies
\[
  |y|_\infty\geq m-\lfloor m/4\rfloor
    \geq\frac{3m}{4}>\frac{3K}{2},
\]
so the square is disjoint from the support of
$\mathcal S_{\Lambda_K,\ell,s}(A)$. Thus \eqref{eq:sampler-locality} and
contractivity give
\[
  \begin{aligned}
    &\norm{(\Phi_{x,\Lambda_J}-\mathrm{id})
      \mathcal S_{\Lambda_K,\ell,s}(A)}\\
    &\quad=\norm{\bigl(\Phi_{x,\Lambda_J}
      -\Phi_{x,\Lambda_J\cap(x+\Lambda_{\lfloor m/4\rfloor})}\bigr)
      \mathcal S_{\Lambda_K,\ell,s}(A)}\\
    &\quad\leq b\e^{-\mu\lfloor m/4\rfloor}\norm{A}.
  \end{aligned}
\]
This error decreases exponentially with the distance $m$ of the
center. There are only $8m$ sites with $|x|_\infty=m$, so the
errors are summable even as $J\to\infty$. Extending the sum to
all $m>2K$ and using $\lfloor m/4\rfloor\geq m/4-1$ gives
\[
  8b\norm{A}\sum_{m=2K+1}^{\infty}
    m\e^{-\mu\lfloor m/4\rfloor}
    \leq CK\e^{-\mu K/2}\norm{A}.
\]

Using contractivity of $\mathcal T_{\Lambda_J,t-s}$ and
$|\Lambda_{2K}|\leq25K^2$, integration gives
\begin{equation}\label{eq:sampler-boundary-comparison}
  \begin{aligned}
    &\norm{\mathcal T_{\Lambda_J,t}(A)-\mathcal S_{\Lambda_K,\ell,t}(A)}\\
    &\quad\leq C_Lt\left[
      K^2\e^{-\mu\ell}
      +K^2\exp\!\left(C\ell^4t-\frac{K}{4\ell}\right)
      +K\e^{-\mu K/2}\right]\norm{A},
        \qquad J\geq2K.
  \end{aligned}
\end{equation}
Here and below, $C_L$ may depend on $L$, but is independent of
$h$, $A$, $J$, $K$, $\ell$, and $t$.

Set $\ell=\lfloor K^{1/8}\rfloor$. For $0\leq t\leq K^{1/8}$
and large $K$, we have $\ell\geq K^{1/8}/2$, and the last term in
\eqref{eq:sampler-boundary-comparison} is bounded by the first. Thus
\begin{equation}\label{eq:sampler-common-approximation}
  \begin{aligned}
    &\sup_{J\geq2K}
      \norm{\mathcal T_{\Lambda_J,t}(A)-\mathcal S_{\Lambda_K,\ell,t}(A)}\\
    &\quad\leq C_LtK^2\left[
      2\e^{-\mu K^{1/8}/2}
      +\exp\!\left(CK^{5/8}-\frac14K^{7/8}\right)\right]\norm{A}\\
    &\quad\leq C_Lt\e^{-cK^{1/8}}\norm{A}.
  \end{aligned}
\end{equation}
Here $c>0$ is independent of $h$. The last step uses that
$K^{7/8}$ grows faster than $K^{5/8}$ and exponential decay absorbs $K^2$.

For $J,J'\geq2K$, comparison with the same
$\mathcal S_{\Lambda_K,\ell,t}(A)$ and the triangle inequality give
\[
  \norm{\mathcal T_{\Lambda_J,t}(A)-\mathcal T_{\Lambda_{J'},t}(A)}
    \leq2C_Lt\e^{-cK^{1/8}}\norm{A}.
\]
Thus the finite-volume evolutions are Cauchy uniformly on bounded
time intervals. Their norm limit
\[
  \mathcal T_t(A)=\lim_{K\to\infty}\mathcal T_{\Lambda_K,t}(A)
\]
exists for every local $A$. For $A\in\mA_{\Lambda_L}$, taking
$J'=2K$ and letting $J\to\infty$ in this Cauchy bound gives
\begin{equation}\label{eq:sampler-volume-limit}
  \norm{\mathcal T_t(A)-\mathcal T_{\Lambda_{2K},t}(A)}
    \leq2C_Lt\e^{-cK^{1/8}}\norm{A},
    \qquad 0\leq t\leq K^{1/8},
\end{equation}
for all sufficiently large integers $K\geq4L$.

Contractivity extends the limit to every $A\in\mA$, still
uniformly on compact time intervals. The maps $\mathcal T_t$ are
unital and completely positive, since these properties pass to
the limit. To pass the semigroup identity to the limit, use
\[
  \begin{aligned}
    &\norm{\mathcal T_{\Lambda_K,t}\mathcal T_{\Lambda_K,s}(A)
            -\mathcal T_t\mathcal T_s(A)}\\
    &\quad\leq
      \norm{\mathcal T_{\Lambda_K,s}(A)-\mathcal T_s(A)}
      +\norm{(\mathcal T_{\Lambda_K,t}-\mathcal T_t)
                \mathcal T_s(A)}\longrightarrow0.
  \end{aligned}
\]
Hence $\mathcal T_{t+s}=\mathcal T_t\mathcal T_s$ and
$\mathcal T_0=\mathrm{id}$. Uniform convergence on compact time
intervals also gives norm continuity of $t\mapsto\mathcal T_t(A)$
for every $A\in\mA$.

We now show that the sampler converges to the expectation in a
KMS state. For each integer $n\geq1$, extend the Gibbs state on
$\Lambda_{2n^8}$ by the normalized trace outside that box.
By weak$^*$ compactness,
choose a subsequence $n_j\to\infty$ along which these states
converge to a state $\omega$. Since the interaction is finite
range, $\omega$ is a $\beta$-KMS state by
\cite[Proposition~6.2.15]{BratteliRobinson1997}.

Fix a self-adjoint $A\in\mA_{\Lambda_L}$. For sufficiently large
$j$, the triangle inequality, \eqref{eq:sampler-volume-limit}
with $K=n_j^8$, and \eqref{eq:sampler-mixing} at time $t=n_j$ give
\[
  \begin{aligned}
    &\norm{\mathcal T_{n_j}(A)-\omega(A)\mathbb1}\\
    &\quad\leq\norm{\mathcal T_{n_j}(A)
      -\mathcal T_{\Lambda_{2n_j^8},n_j}(A)}\\
    &\qquad+\norm{\mathcal T_{\Lambda_{2n_j^8},n_j}(A)
      -\tr_{\Lambda_{2n_j^8}}(\gamma_{\Lambda_{2n_j^8}}A)\mathbb1}\\
    &\qquad+\bigl|\tr_{\Lambda_{2n_j^8}}
      (\gamma_{\Lambda_{2n_j^8}}A)-\omega(A)\bigr|\\
    &\quad\leq\left[2C_Ln_j\e^{-cn_j}
      +2|\Lambda_{2n_j^8}|\e^{-\kappa n_j}\right]\norm{A}\\
    &\qquad+\bigl|\tr_{\Lambda_{2n_j^8}}
      (\gamma_{\Lambda_{2n_j^8}}A)-\omega(A)\bigr|.
  \end{aligned}
\]
The right-hand side tends to zero: the first two terms vanish
since $|\Lambda_{2n_j^8}|\leq25n_j^{16}$, and the last term
vanishes by the choice of $\omega$. For every $t\geq n_j$,
the semigroup property, unitality, and contractivity imply
\[
  \begin{aligned}
    \norm{\mathcal T_t(A)-\omega(A)\mathbb1}
      &=\norm{\mathcal T_{t-n_j}
        \bigl(\mathcal T_{n_j}(A)-\omega(A)\mathbb1\bigr)}\\
      &\leq\norm{\mathcal T_{n_j}(A)-\omega(A)\mathbb1}.
  \end{aligned}
\]
This proves convergence as $t\to\infty$. Linearity extends
it to all local observables, and contractivity and norm density
extend it to every $A\in\mA$. Thus
\[
  \lim_{t\to\infty}\norm{\mathcal T_t(A)-\omega(A)\mathbb1}=0,
    \qquad A\in\mA.
\]

\paragraph*{3. Invariance of KMS states.}
Let $\varphi$ be any $\beta$-KMS state. We first prove that it is
invariant under each limiting update $\Phi_x$.
Write $\tau_{Z,t}(A)=\exp(\i tH_Z)A\exp(-\i tH_Z)$ for the
physical evolution, with infinite-volume limit $\tau_t$.
The balance identities \eqref{eq:sampler-balance} give
\begin{equation}\label{eq:sampler-finite-shifts}
  \begin{aligned}
    \tau_{Z,-\i\beta/2}(L_{x,a,Z})
      &=Q_Z^{-1}L_{x,a,Z}Q_Z=L_{x,a,Z}^*,\\
    \tau_{Z,\i\beta/2}(D_{x,Z})
      &=Q_ZD_{x,Z}Q_Z^{-1}=D_{x,Z}^*.
  \end{aligned}
\end{equation}
To pass to the limit at imaginary times, consider the bounded
analytic function $z\mapsto\tau_{Z,z}(L_{x,a,Z})$ on the strip
$-\beta/2\leq\operatorname{Im}z\leq0$. By
\eqref{eq:sampler-finite-shifts}, its values on the upper and lower
boundary lines are $\tau_{Z,t}(L_{x,a,Z})$ and
$\tau_{Z,t}(L_{x,a,Z}^*)$, respectively. Both converge in norm
on bounded real-time intervals and are uniformly bounded.

The three-lines theorem bounds the norm of a bounded analytic
function on a strip by its boundary norms
\cite[Proposition~5.3.5]{BratteliRobinson1997}. Apply it to
$\e^{-z^2}$ times the difference between two volumes.
This factor decays as $|t|\to\infty$ on both boundary lines,
making convergence uniform there. The theorem then gives uniform norm convergence on bounded
parts of the closed strip, extending $\tau_t(L_{x,a})$ to the
lower boundary. For $D_{x,Z}$, the same argument uses the strip
$0\leq\operatorname{Im}z\leq\beta/2$. Hence
\begin{equation}\label{eq:sampler-half-strip}
  \tau_{-\i\beta/2}(L_{x,a})=L_{x,a}^*,
  \qquad
  \tau_{\i\beta/2}(D_x)=D_x^*.
\end{equation}

We use the halfway form of the KMS identity
\cite[Propositions~5.3.3 and~5.3.7]{BratteliRobinson1997},
\[
  \varphi(BC)=\varphi\bigl(\tau_{-\i\beta/2}(C)
                           \tau_{\i\beta/2}(B)\bigr).
\]
It applies when the evolutions of $B$ and $C$ are analytic inside
the strips $0<\operatorname{Im}z<\beta/2$ and
$-\beta/2<\operatorname{Im}z<0$, respectively, and extend
continuously in norm to their boundary lines. We have just shown
this for $D_x$ on the upper strip and $L_{x,a}$ on the lower
strip. Taking adjoints gives the corresponding extensions for
$L_{x,a}^*$ and $D_x^*$. Choose $A$ whose evolution is analytic
for all complex times.
Using \eqref{eq:sampler-half-strip} gives
\begin{align*}
  \varphi(L_{x,a}^*AL_{x,a})
    &=\varphi\bigl(\tau_{-\i\beta/2}(L_{x,a})
                    \tau_{\i\beta/2}(L_{x,a}^*A)\bigr)
      =\varphi\bigl(L_{x,a}^*L_{x,a}\tau_{\i\beta/2}(A)\bigr),\\
  \varphi(D_xA)
    &=\varphi\bigl(\tau_{\i\beta/2}(D_x)
                    \tau_{\i\beta/2}(A)\bigr)
      =\varphi\bigl(D_x^*\tau_{\i\beta/2}(A)\bigr),\\
  \varphi(AD_x^*)
    &=\varphi\bigl(\tau_{-\i\beta/2}(D_x^*)
                    \tau_{\i\beta/2}(A)\bigr)
      =\varphi\bigl(D_x\tau_{\i\beta/2}(A)\bigr).
\end{align*}
The middle line follows from invariance of $\varphi$ under real-time evolution
and analytic continuation to the upper half-strip.
Summing the first identity over $a$, adding the last two, and
using the balance identity in \eqref{eq:sampler-limiting-identities} gives
\[
  \varphi(\mathcal K_x(A))
    =\varphi\!\left(
      \left[\sum_{a=1}^9L_{x,a}^*L_{x,a}+D_x+D_x^*\right]
        \tau_{\i\beta/2}(A)\right)=0.
\]
Such observables are norm dense and $\mathcal K_x$ is bounded,
so $\varphi(\mathcal K_x(A))=0$ holds for every
$A\in\mA$. Applying this also to $\mathcal K_x^{n-1}(A)$, the
exponential series yields
\[
  \varphi(\Phi_x(A))=\varphi(A),\qquad A\in\mA.
\]

We now pass from individual updates to the full evolution. For
local $A\in\mA_{\Lambda_L}$, differentiation gives
\[
  \begin{aligned}
    &\varphi(\mathcal S_{\Lambda_K,\ell,t}(A))-\varphi(A)\\
    &\quad=\int_0^t\sum_{x\in\Lambda_K}
      \varphi\bigl((\Phi_{x,x+\Lambda_\ell}-\Phi_x)
        \mathcal S_{\Lambda_K,\ell,s}(A)\bigr)\,\D s.
  \end{aligned}
\]
Contractivity and the locality estimate in
\eqref{eq:sampler-limiting-identities} imply
\[
  |\varphi(\mathcal S_{\Lambda_K,\ell,t}(A))-\varphi(A)|
    \leq bt|\Lambda_K|\e^{-\mu\ell}\norm{A}.
\]
Fix $t$ and let $K\to\infty$, with $\ell=\lfloor K^{1/8}\rfloor$
as above. The right-hand side tends to zero. Letting $J\to\infty$
in \eqref{eq:sampler-common-approximation} also shows that
$\mathcal S_{\Lambda_K,\ell,t}(A)\to\mathcal T_t(A)$ in norm.
Thus, first for local $A$ and then by norm density,
\[
  \varphi(\mathcal T_t(A))=\varphi(A),
    \qquad A\in\mA,\quad t\geq0.
\]

\paragraph*{4. Conclusion.}
For every $\beta$-KMS state $\varphi$, parts~2 and~3 give
\[
  \varphi(A)=\varphi(\mathcal T_t(A))
    \longrightarrow\omega(A),\qquad A\in\mA.
\]
Thus $\omega$ is the unique KMS state. Every translate of $\omega$
is again KMS, since the Hamiltonian is translation invariant.
Uniqueness therefore makes $\omega$ translation invariant.
Every weak$^*$ limit point of the finite-volume Gibbs states is
KMS by \cite[Proposition~6.2.15]{BratteliRobinson1997}, and hence
equals $\omega$. Weak$^*$ compactness then gives local convergence
of the full sequence to $\omega$.

\section{Proof of \texorpdfstring{\hyperref[lem:probability-return]{Lemma~\ref*{lem:probability-return}}}{Lemma~\ref*{lem:probability-return}}}\label{sec:probability-proof}

For $A\in\mA_Z$, we write $A_{\eta,\xi}$ for its matrix entries in the
occupation basis, indexed by $\eta,\xi\in\{0,1,2\}^Z$.

Fix $L\geq1$ and take $K>L$, so that
$\Lambda_L^{+}\subset\Lambda_K$. By
Lemma~\ref{lem:kms-uniqueness}, the finite-volume Gibbs states converge
locally to the fixed state $\omega$. In particular,
\begin{equation}\label{eq:empty-probability-limit}
  \omega(\nu_{\Lambda_L})
    =\lim_{K\to\infty}
      \frac{\tr_{\Lambda_K}
        \bigl[\nu_{\Lambda_L}\exp(-\beta H_{\Lambda_K})\bigr]}
           {\mathcal Z_{\Lambda_K}}.
\end{equation}
We estimate the numerator and denominator separately, choosing
the comparisons so that the same exterior partition function
$\mathcal Z_{\Lambda_K\setminus \Lambda_L^{+}}$ appears in both.

\paragraph*{1. Lower bound on the numerator.}
  For the numerator, expand the trace in the occupation basis.
  Every diagonal entry of
  $\exp(-\beta H_{\Lambda_K})$ is positive, so retaining only the
  configurations with occupation two on $B_L$ gives
  \begin{align}
    \tr_{\Lambda_K}
       \bigl[\nu_{\Lambda_L}\exp(-\beta H_{\Lambda_K})\bigr]
    &=\sum_{\substack{\eta\in\{0,1,2\}^{\Lambda_K}\\
                      \eta|_{\Lambda_L}=0}}
       \bigl[\exp(-\beta H_{\Lambda_K})\bigr]_{\eta,\eta} \notag\\[0.5em]
    &\geq
      \sum_{\substack{\eta\in\{0,1,2\}^{\Lambda_K}\\
                      \eta|_{\Lambda_L}=0,\ \eta|_{B_L}=2}}
       \bigl[\exp(-\beta H_{\Lambda_K})\bigr]_{\eta,\eta}.
      \label{eq:retain-seeded-boundary}
  \end{align}
  Split the Hamiltonian at the boundary of $\Lambda_L^{+}$:
  \begin{equation}\label{eq:outer-Hamiltonian-splitting}
    H_{\Lambda_K}
      =\underbrace{H_{\Lambda_L^{+}}}_{\text{inside}}
       +\underbrace{H_{\Lambda_K\setminus\Lambda_L^{+}}}_{\text{outside}}
       -\underbrace{
         \sum_{\substack{x\to y,\ x,y\in\Lambda_K\\
                   |\{x,y\}\cap\Lambda_L^{+}|=1}}F_{xy}
         }_{\text{bonds crossing the cut}}.
  \end{equation}
  All bond matrices have nonnegative entries in the occupation basis.
  Writing $\leq_{\mathrm{ent}}$ for entrywise comparison, we have
  \[
    0\leq_{\mathrm{ent}}
      2h|\Lambda_K|\mathbb{1}
        -H_{\Lambda_L^{+}}-H_{\Lambda_K\setminus\Lambda_L^{+}}
      \leq_{\mathrm{ent}}
      2h|\Lambda_K|\mathbb{1}-H_{\Lambda_K}.
  \]
  Taking powers and summing the exponential series gives, after
  multiplying by $\e^{-2\beta h|\Lambda_K|}$,
  \[
    \exp(-\beta H_{\Lambda_K})
      \geq_{\mathrm{ent}}
      \exp\!\left[-\beta\left(
        H_{\Lambda_L^{+}}+H_{\Lambda_K\setminus\Lambda_L^{+}}
      \right)\right]
      =\exp(-\beta H_{\Lambda_L^{+}})
       \otimes\exp(-\beta H_{\Lambda_K\setminus\Lambda_L^{+}}).
  \]
  Substituting into \eqref{eq:retain-seeded-boundary}, the trace factors:
  \begin{align}
    \tr_{\Lambda_K}
      \bigl[\nu_{\Lambda_L}\exp(-\beta H_{\Lambda_K})\bigr]
    &\geq
      \scal{u_{\Lambda_L^{+}}}
        {\exp(-\beta H_{\Lambda_L^{+}})u_{\Lambda_L^{+}}}
      \;\tr_{\Lambda_K\setminus\Lambda_L^{+}}
        \exp(-\beta H_{\Lambda_K\setminus\Lambda_L^{+}})
      \notag\\[0.5em]
    &=G_{\Lambda_L^{+}}\mathcal Z_{\Lambda_K\setminus\Lambda_L^{+}}.
      \label{eq:numerator-return-bound}
  \end{align}

\paragraph*{2. Upper bound on the denominator.}
  For the denominator, use the Golden--Thompson inequality
  \cite[Eq.~(1.1)]{ForresterThompson2014}:
  \[
    \tr\exp(A+B)\leq\tr\bigl[\exp(A)\exp(B)\bigr]
       \leq\e^{\norm{B}}\tr\exp(A),
       \qquad A=A^*,\quad B=B^*.
  \]
  In \eqref{eq:outer-Hamiltonian-splitting}, at most $4(2L+2)$ bonds
  cross the cut, and $\norm{F_{xy}}=1$. Hence
  \begin{align}
    \mathcal Z_{\Lambda_K}
      &\leq\e^{4\beta(2L+2)}
        \tr_{\Lambda_K}
        \left[
          \exp(-\beta H_{\Lambda_L^{+}})
          \otimes\exp(-\beta H_{\Lambda_K\setminus\Lambda_L^{+}})
        \right] \notag\\[0.4em]
      &=\e^{4\beta(2L+2)}
        \mathcal Z_{\Lambda_L^{+}}
        \mathcal Z_{\Lambda_K\setminus\Lambda_L^{+}}.
        \label{eq:outer-partition-bound}
  \end{align}
  Next, separate the sites of $B_L$ from each other and from $\Lambda_L$:
  \[
    H_{\Lambda_L^{+}}
      =\underbrace{H_{\Lambda_L}}_{\text{core}}
       +\underbrace{h\sum_{x\in B_L}(2-n_x)}_{\text{individual boundary sites}}
       -\underbrace{
         \sum_{\substack{x\to y,\ x,y\in\Lambda_L^{+}\\
                    \{x,y\}\cap B_L\ne\varnothing}}F_{xy}
         }_{4(2L+1)\text{ bonds}}.
  \]
  The same trace inequality gives
  \begin{align}
    \mathcal Z_{\Lambda_L^{+}}
      &\leq\e^{4\beta(2L+1)}\mathcal Z_{\Lambda_L}
         \tr_{B_L}\exp\!\left[-\beta h\sum_{x\in B_L}(2-n_x)\right]
         \notag\\[0.5em]
      &=\e^{4\beta(2L+1)}\mathcal Z_{\Lambda_L}
         \prod_{x\in B_L}\tr_{\{x\}}\exp[-\beta h(2-n_x)]
         \notag\\[0.5em]
      &=\e^{4\beta(2L+1)}\mathcal Z_{\Lambda_L}
         (1+\e^{-\beta h}+\e^{-2\beta h})^{|B_L|}
         \leq\e^{4\beta(2L+1)}\mathcal Z_{\Lambda_L}\,3^{|B_L|},
        \label{eq:inner-partition-bound}
  \end{align}
  where the last inequality uses $h\geq0$. Combining
  \eqref{eq:outer-partition-bound} and \eqref{eq:inner-partition-bound},
  \begin{equation}\label{eq:denominator-bound}
    \mathcal Z_{\Lambda_K}
    \leq\e^{4\beta(2L+2)+4\beta(2L+1)}\,3^{|B_L|}
       \mathcal Z_{\Lambda_L}\mathcal Z_{\Lambda_K\setminus\Lambda_L^{+}}
    =\e^{(4\beta+\log3)|B_L|}
       \mathcal Z_{\Lambda_L}\mathcal Z_{\Lambda_K\setminus\Lambda_L^{+}}.
  \end{equation}

\paragraph*{3. Infinite-volume limit.}
  Combining \eqref{eq:numerator-return-bound} and
  \eqref{eq:denominator-bound} cancels
  $\mathcal Z_{\Lambda_K\setminus \Lambda_L^{+}}$ and gives
  \[
    \frac{\tr_{\Lambda_K}
      \bigl[\nu_{\Lambda_L}\exp(-\beta H_{\Lambda_K})\bigr]}
         {\mathcal Z_{\Lambda_K}}
      \geq\e^{-(4\beta+\log3)|B_L|}
          \frac{G_{\Lambda_L^{+}}}{\mathcal Z_{\Lambda_L}}.
  \]
  The right-hand side is independent of $K$. Thus we may take the
  local limit in \eqref{eq:empty-probability-limit}, keeping $L$
  fixed, to obtain \eqref{eq:probability-return-bound}.

\section{Proof of \texorpdfstring{\hyperref[lem:return-amplitude]{Lemma~\ref*{lem:return-amplitude}}}{Lemma~\ref*{lem:return-amplitude}}}\label{sec:return-proof}

The proof prepares four smaller squares by filling one row and two
column segments. A comparison of return amplitudes then reduces the
estimate to a scalar recursion. This repeated subdivision is based on the multiscale branching idea of \cite[Section~6]{Bouch15}.

\paragraph*{1. Setup of the recursive estimate.}
  Fix $\beta>0$ and $h\geq0$. For a finite lattice subset $Z$, set
  \[
    A_Z=2h|Z|\mathbb1-H_Z
      =h\sum_{x\in Z}n_x
        +\sum_{\substack{x\to y\\x,y\in Z}}F_{xy}.
  \]
  This matrix is self-adjoint and has nonnegative entries in the
  occupation basis. Moreover,
  \begin{equation}\label{eq:return-shift}
    \exp(-\beta H_Z)
      =\e^{-2\beta h|Z|}\exp(\beta A_Z).
  \end{equation}

  We will use operations on the configuration that can lower boundary occupations, so we need a class of
  initial vectors that is preserved by our subdivision procedure. Let
  $\mathcal U_{\Lambda_L^{+}}$ be the set of occupation-basis vectors
  on $\Lambda_L^{+}$ that are empty on $\Lambda_L$, have occupation one or two on
  $B_L\setminus\{(-L-1,-L-1)\}$, and have arbitrary occupation at the
  remaining corner. For $L\geq0$, set
  \begin{equation}\label{eq:minimal-shifted-return}
    g_{\Lambda_L^{+}}
      =\min_{u\in\mathcal U_{\Lambda_L^{+}}}
         \scal{u}{\exp(\beta A_{\Lambda_L^{+}})u}.
  \end{equation}
  The exponential series gives $g_{\Lambda_L^{+}}\geq1$.
  Since $u_{\Lambda_L^{+}}\in\mathcal U_{\Lambda_L^{+}}$,
  \eqref{eq:return-shift} implies
  \begin{equation}\label{eq:return-field-comparison}
    G_{\Lambda_L^{+}}\geq\e^{-2\beta h|\Lambda_L^{+}|}g_{\Lambda_L^{+}}.
  \end{equation}

\paragraph*{2. Comparison along a sequence of conversions.}
  Let
  \[
    u=\bigotimes_{z\in Z}|a_z\rangle,\qquad
    v=\bigotimes_{z\in Z}|b_z\rangle,
    \qquad a_z,b_z\in\{0,1,2\},
  \]
  be two occupation-basis vectors on $Z$. We say that $u$ and $v$ are
  connected by an allowed move, called a conversion, if there is a
  directed bond $x\to y$ in $Z$ such that $a_z=b_z$ for every
  $z\in Z\setminus\{x,y\}$ and one of the following four cases holds:
  \[
    \begin{aligned}
      (a_x,a_y)&=(1,0), & (b_x,b_y)&=(0,2),\\
      (a_x,a_y)&=(0,2), & (b_x,b_y)&=(1,0),\\
      (a_x,a_y)&=(2,0), & (b_x,b_y)&=(1,2),\\
      (a_x,a_y)&=(1,2), & (b_x,b_y)&=(2,0).
    \end{aligned}
  \]
  A conversion corresponds to one particle hopping along the bond from
  $x$ to $y$ and splitting into two, or two particles hopping in the
  opposite direction and merging into one.
  For the bond that realizes this conversion, $F_{xy}u=v$, so the
  corresponding matrix entry is one.
  Now suppose that $u$ and $v$ are joined by a sequence
  $u=u_0,u_1,\ldots,u_q=v$ of $q\geq1$ allowed conversions.
  For each $j=1,\ldots,q$, choose a bond $x\to y$ with
  $F_{xy}u_{j-1}=u_j$. The matrix $A_Z-F_{xy}$ is the sum of the
  remaining field and bond terms, all of which have nonnegative entries.
  Its action on the basis vector $u_{j-1}$ therefore has nonnegative
  coefficients, giving
  \[
    A_Zu_{j-1}=u_j+(A_Z-F_{xy})u_{j-1}\geq_{\mathrm{ent}}u_j,
  \]
  where $\geq_{\mathrm{ent}}$ compares the coefficients in the
  occupation basis. Multiplication by $A_Z$ preserves these comparisons,
  so iteration gives
  \[
    A_Z^q u\geq_{\mathrm{ent}}A_Z^{q-1}u_1
      \geq_{\mathrm{ent}}\cdots\geq_{\mathrm{ent}}u_q=v.
  \]
  The exponential series shows that $\exp(\beta A_Z/2)$ also has
  nonnegative entries and therefore preserves this comparison.
  Taking norms and commuting $A_Z^q$ with the exponential gives
  \begin{align*}
    \norm{\exp(\beta A_Z/2)v}
      &\leq\norm{\exp(\beta A_Z/2)A_Z^q u}\\
      &=\norm{A_Z^q\exp(\beta A_Z/2)u}\\
      &\leq\norm{A_Z}^q\norm{\exp(\beta A_Z/2)u}.
  \end{align*}
  Self-adjointness of $A_Z$ gives
  \begin{align*}
    \norm{\exp(\beta A_Z/2)u}^2
      &=\scal{u}{\exp(\beta A_Z/2)^*\exp(\beta A_Z/2)u}\\
      &=\scal{u}{\exp(\beta A_Z)u},
  \end{align*}
  and the same identity holds for $v$. Squaring the preceding norm
  inequality and dividing by $\norm{A_Z}^{2q}$ therefore proves that
  any two occupation-basis vectors $u$ and $v$ on $Z$ connected by a
  sequence of $q\geq1$ allowed conversions satisfy
  \begin{equation}\label{eq:return-conversion-comparison}
    \scal{u}{\exp(\beta A_Z)u}
      \geq\norm{A_Z}^{-2q}\scal{v}{\exp(\beta A_Z)v}.
  \end{equation}

\begin{figure}[htbp]
    \centering
    \begin{tikzpicture}[x=0.90cm,y=0.90cm]
      \foreach \a/\b in {-4/-4,0/-4,-4/0,0/0} {
        \fill[blue!8] (\a-.35,\b-.35)
          rectangle (\a+3.35,\b+3.35);
        \draw[blue!65!black,thick]
          (\a,\b+3)--(\a,\b)--(\a+3,\b);
      }
      \foreach \x in {-4,...,3} {
        \foreach \y in {-4,...,3} {
          \fill[black!25] (\x,\y) circle[radius=1.3pt];
        }
      }
      \draw[orange!80!black,very thick,->] (-4,0)--(3,0);
      \draw[orange!80!black,very thick,->] (0,-4)--(0,-1);
      \draw[orange!80!black,very thick,->] (0,0)--(0,3);
      \foreach \x/\y in {-4/0,0/-4,0/0} {
        \draw[fill=white] (\x,\y) circle[radius=2.6pt];
      }
      \node[anchor=south,font=\small,fill=blue!8,inner sep=1pt]
        at (-2,.15) {(i)};
      \node[anchor=west,font=\small,fill=blue!8,inner sep=1pt]
        at (.15,-2) {(ii)};
      \node[anchor=west,font=\small,fill=blue!8,inner sep=1pt]
        at (.15,2) {(iii)};
    \end{tikzpicture}
    \caption{The three sweeps for $L_k=3$. Each shaded square is a
      translate of $\Lambda_{L_{k-1}}^{+}$, with its added bottom
      and left edges marked. The circled sources become corners
      of the smaller squares, where arbitrary occupation is allowed.
      The central circled site is the origin, where sweep (iii) starts.}
    \label{fig:return-subdivision}
  \end{figure}
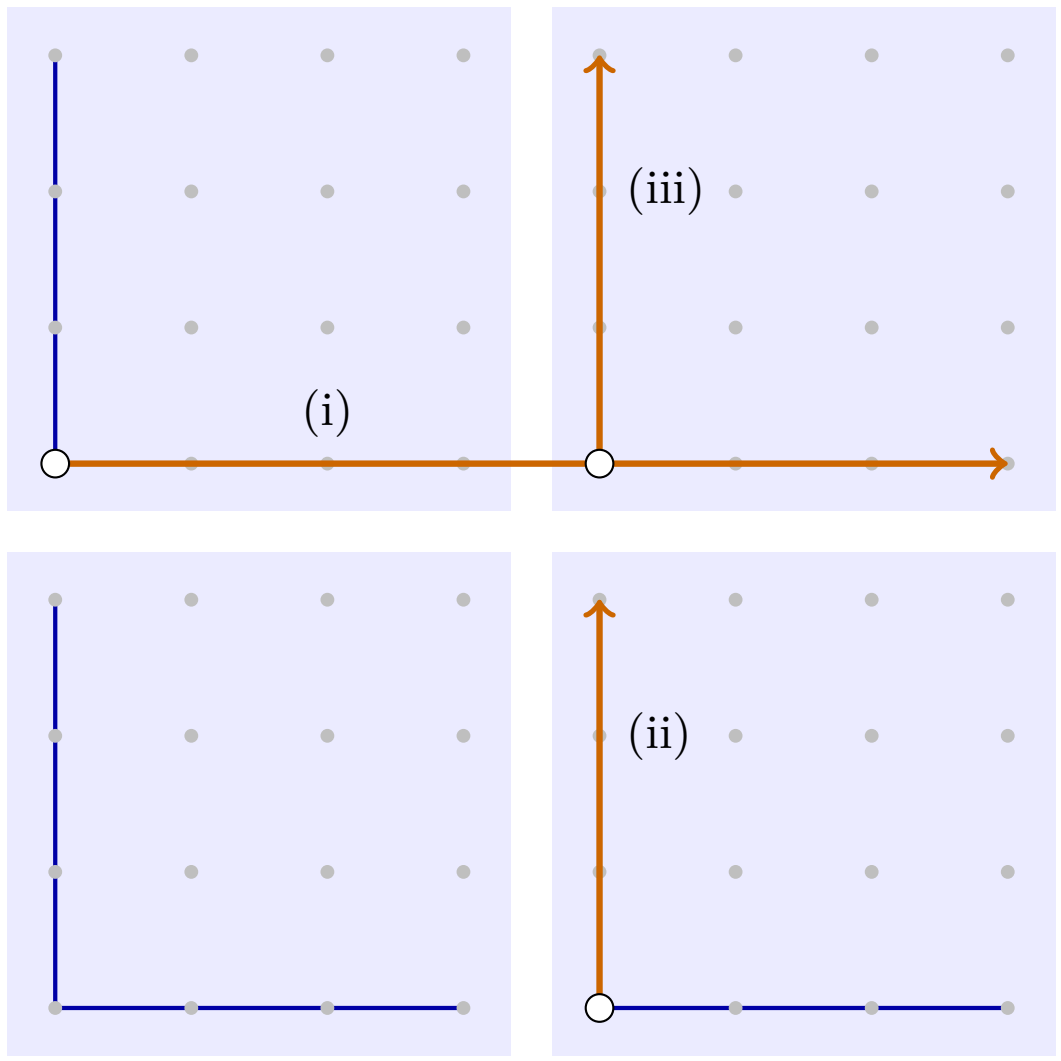

\paragraph*{3. Subdivision into four squares.}
  Recall that $L_k=2^k-1$.
  Fix $k\geq1$ and $u\in\mathcal U_{\Lambda_{L_k}^{+}}$.
  We construct an occupation-basis vector $v$ from $u$ using the
  allowed conversions $(1,0)\mapsto(0,2)$ and $(2,0)\mapsto(1,2)$.
  In each of the following three stages, apply these conversions
  successively along the indicated row or column, starting at the
  specified source; see Figure~\ref{fig:return-subdivision}.
  \begin{enumerate}[label=(\roman*),nosep]
    \item Fill the row $(x_1,0)$, $x_1=-L_k,\ldots,L_k$, from
      left to right, starting with the source $(-L_k-1,0)$.
    \item Fill the lower column $(0,x_2)$, $x_2=-L_k,\ldots,-1$,
      from bottom to top, starting with the source $(0,-L_k-1)$.
    \item Fill the upper column $(0,x_2)$, $x_2=1,\ldots,L_k$,
      from bottom to top, starting with the source $(0,0)$.
  \end{enumerate}

  The sources in (i) and (ii) have occupation one or two by the
  definition of $\mathcal U_{\Lambda_{L_k}^{+}}$. Each target is
  empty when first reached and receives occupation two, so it can
  supply the next site. The source $(0,0)$ in (iii) has occupation
  one after the horizontal sweep and can therefore supply the upper
  column segment. Thus every step is an allowed conversion. Their
  total number is
  \[
    q_k=\underbrace{(2L_k+1)}_{\text{row}}
      +\underbrace{L_k}_{\text{lower column}}
      +\underbrace{L_k}_{\text{upper column}}
      =2^{k+2}-3.
  \]
  The enlarged square is the disjoint union
  \[
    \Lambda_{L_k}^{+}
      =\bigsqcup_{z\in\{-2^{k-1},2^{k-1}\}^2}
         \bigl(z+\Lambda_{L_{k-1}}^{+}\bigr).
  \]
  On each of these four squares, $v$ is empty on the translated
  core and has occupation one or two on its added bottom and left
  edges, except possibly at their common corner. The three sources
  $(-L_k-1,0)$, $(0,-L_k-1)$, and $(0,0)$ that may become empty
  are the bottom-left corners of three smaller squares; the fourth
  inherits the original bottom-left corner. Thus the tensor factor
  $v_z$ of $v$
  on $z+\Lambda_{L_{k-1}}^{+}$ is a translate of a vector in
  $\mathcal U_{\Lambda_{L_{k-1}}^{+}}$.

  Deleting the bonds between the four squares gives
  \[
    0\leq_{\mathrm{ent}}
      \sum_{z\in\{-2^{k-1},2^{k-1}\}^2}
        A_{z+\Lambda_{L_{k-1}}^{+}}
      \leq_{\mathrm{ent}}A_{\Lambda_{L_k}^{+}}.
  \]
  The exponential of the sum factors over the four squares.
  Comparing exponential series therefore yields
  \begin{align}
    \scal{v}{\exp(\beta A_{\Lambda_{L_k}^{+}})v}
      &\geq\prod_{z\in\{-2^{k-1},2^{k-1}\}^2}
        \scal{v_z}{\exp(\beta A_{z+\Lambda_{L_{k-1}}^{+}})v_z}
        \notag\\[0.5em]
      &\geq g_{\Lambda_{L_{k-1}}^{+}}^{\,4}.
      \label{eq:four-square-return}
  \end{align}
  The last inequality uses translation invariance and
  \eqref{eq:minimal-shifted-return}.

\paragraph*{4. Iteration of the return bound.}
  Counting the bonds in $\Lambda_L^{+}$ and using $\norm{F_{xy}}=1$
  and $\norm{n_x}=2$ gives
  \[
    \norm{A_{\Lambda_L^{+}}}
      \leq2h|\Lambda_L^{+}|+2(2L+2)(2L+1)
      \leq2(1+h)|\Lambda_L^{+}|.
  \]
  For each $u\in\mathcal U_{\Lambda_{L_k}^{+}}$, the vector $v$
  constructed in Part~3 is connected to $u$ by $q_k$ allowed
  conversions within $\Lambda_{L_k}^{+}$. Applying
  \eqref{eq:return-conversion-comparison} to these two vectors gives
  \[
    \scal{u}{\exp(\beta A_{\Lambda_{L_k}^{+}})u}
      \geq\norm{A_{\Lambda_{L_k}^{+}}}^{-2q_k}
        \scal{v}{\exp(\beta A_{\Lambda_{L_k}^{+}})v}.
  \]
  Using the norm bound above and \eqref{eq:four-square-return}, then
  taking the minimum over $u\in\mathcal U_{\Lambda_{L_k}^{+}}$,
  gives the recursion
  \begin{equation}\label{eq:shifted-return-recursion}
    g_{\Lambda_{L_k}^{+}}
      \geq\bigl(2(1+h)|\Lambda_{L_k}^{+}|\bigr)^{-2q_k}
        g_{\Lambda_{L_{k-1}}^{+}}^{\,4}.
  \end{equation}
  For $L_0=0$, every $u\in\mathcal U_{\Lambda_0^{+}}$ has occupation
  at least one at both $(-1,0)$ and $(0,-1)$. Discarding the bond
  terms and comparing exponential series gives
  \[
    g_{\Lambda_{L_0}^{+}}
      \geq\min_{u\in\mathcal U_{\Lambda_0^{+}}}
        \scal{u}{\exp\!\left(\beta h\sum_{x\in\Lambda_0^{+}}n_x\right)u}
      =\e^{2\beta h}.
  \]

  Taking logarithms in \eqref{eq:shifted-return-recursion} and
  dividing by $|\Lambda_{L_k}^{+}|$ gives
  \begin{align*}
    \frac{\log g_{\Lambda_{L_k}^{+}}}{|\Lambda_{L_k}^{+}|}
      &\geq\frac{4\log g_{\Lambda_{L_{k-1}}^{+}}}
                   {|\Lambda_{L_k}^{+}|}
        -\frac{2q_k}{|\Lambda_{L_k}^{+}|}
          \log\bigl(2(1+h)|\Lambda_{L_k}^{+}|\bigr)\\
      &\geq\frac{\log g_{\Lambda_{L_{k-1}}^{+}}}
                   {|\Lambda_{L_{k-1}}^{+}|}
        -\frac{2}{2^k}\log\bigl((1+h)2^{2k+3}\bigr)\\
      &=\frac{\log g_{\Lambda_{L_{k-1}}^{+}}}
                   {|\Lambda_{L_{k-1}}^{+}|}
        -\frac{2}{2^k}\bigl(\log(1+h)+(2k+3)\log2\bigr).
  \end{align*}
  Here we used $q_k\leq2^{k+2}$ and
  $|\Lambda_{L_k}^{+}|=4^{k+1}=4|\Lambda_{L_{k-1}}^{+}|$.
  Iterating from the base case, where $|\Lambda_{L_0}^{+}|=4$, gives
  \begin{align*}
    \frac{\log g_{\Lambda_{L_k}^{+}}}{|\Lambda_{L_k}^{+}|}
      &\geq\frac{\beta h}{2}
        -2\log(1+h)\sum_{j=1}^k2^{-j}
        -2\log2\sum_{j=1}^k(2j+3)2^{-j}\\
      &\geq\frac{\beta h}{2}-2\log(1+h)-14\log2.
  \end{align*}
  Here we used $\sum_{j\geq1}2^{-j}=1$ and
  $\sum_{j\geq1}j2^{-j}=2$.
  Equation~\eqref{eq:return-field-comparison} now proves
  \eqref{eq:return-amplitude-lower-bound}.

\statement{Acknowledgments}

This work was funded by the \foreignlanguage{ngerman}{Deutsche Forschungsgemeinschaft} (DFG, German Research Foundation) – 470903074.

\statement{AI declaration}

Large language models were used to assist with the exploration of models
and proof strategies, the development and checking of mathematical
arguments, literature searches, and the drafting and revision of the
manuscript. The author takes full responsibility for its mathematical
content and references.

\printbibliography

\end{document}